\documentclass[aps, prd, floatfix, nofootinbib, superscriptaddress, twocolumn]{revtex4-1}

\usepackage{latexsym}
\usepackage{amsmath}
\usepackage{amssymb}
\usepackage{amsfonts}

\usepackage[mathscr,scaled=1.15]{urwchancal}
\DeclareFontFamily{OT1}{pzc}{}
\DeclareFontShape{OT1}{pzc}{m}{it}%
{<-> s * [1.15] pzcmi7t}{}
\DeclareMathAlphabet{\mathpzc}{OT1}{pzc}{m}{it}

\usepackage{color}

\usepackage{supertabular}
\usepackage{graphicx}

\definecolor{purple}{rgb}{0.5,0,0.5}
\definecolor{blue}{rgb}{0.0,0,0.9}
\definecolor{prdblue}{rgb}{0.133,0.118,0.498}
\usepackage[colorlinks=true, pdfstartview=FitV, linkcolor=prdblue, citecolor= prdblue, urlcolor=prdblue]{hyperref}

\begin{document}


\title{$\,$\\[-6ex]\hspace*{\fill}{\sf\small{\emph{Preprint no}. NJU-INP 010/19}}\\[1ex]
Elastic electromagnetic form factors of vector mesons}



\author{Y.-Z. Xu}
\email[]{xuyz@smail.nju.edu.cn}
\affiliation{School of Physics, Nanjing University, Nanjing, Jiangsu 210093, China}

\author{D. Binosi}
\email[]{binosi@ectstar.eu}
\affiliation{European Centre for Theoretical Studies in Nuclear Physics
and Related Areas (ECT$^\ast$) and Fondazione Bruno Kessler\\ Villa Tambosi, Strada delle Tabarelle 286, I-38123 Villazzano (TN) Italy}

\author{Z.-F. Cui}
\email[]{phycui@nju.edu.cn}
\affiliation{School of Physics, Nanjing University, Nanjing, Jiangsu 210093, China}
\affiliation{Institute for Nonperturbative Physics, Nanjing University, Nanjing, Jiangsu 210093, China}

\author{B.-L. Li}
\email[]{blli@njnu.edu.cn}
\affiliation{Department of Physics, Nanjing Normal University, Nanjing 210023, China}

\author{C. D. Roberts}
\email[]{cdroberts@nju.edu.cn}
\affiliation{School of Physics, Nanjing University, Nanjing, Jiangsu 210093, China}
\affiliation{Institute for Nonperturbative Physics, Nanjing University, Nanjing, Jiangsu 210093, China}

\author{S.-S.~Xu}
\email[]{xuss@njupt.edu.cn}
\affiliation{College of Science, Nanjing University of Posts and Telecommunications, Nanjing 210023, China}

\author{H.-S. Zong}
\email[]{zonghs@nju.edu.cn}
\affiliation{School of Physics, Nanjing University, Nanjing, Jiangsu 210093, China}

\date{10 November 2019}

\begin{abstract}
A symmetry-preserving approach to the two valence-body continuum bound-state problem is used to calculate the elastic electromagnetic form factors of the $\rho$-meson and subsequently to study the evolution of vector-meson form factors with current-quark mass.  To facilitate a range of additional comparisons, $K^\ast$ form factors are also computed.  The analysis reveals that: vector mesons are larger than pseudoscalar mesons; composite vector mesons are non-spherical, with magnetic and quadrupole moments that deviate $\sim 30$\% from point-particle values; in many ways, vector-meson properties are as much influenced by emergent mass as those of pseudoscalars; and vector meson electric form factors possess a zero at spacelike momentum transfer.  Qualitative similarities between the electric form factors of the $\rho$ and the proton, $G_E^p$, are used to argue that the character of emergent mass in the Standard Model can force a zero in $G_E^p$.  Morover, the existence of a zero in vector meson electric form factors entails that a single-pole vector meson dominance model can only be of limited use in estimating properties of off-shell vector mesons, providing poor guidance for systems in which the Higgs-mechanism of mass generation is dominant.
\end{abstract}


\maketitle


\section{Introduction}\label{introduction}
%
The Lagrangian that defines quantum chromodynamics (QCD) appears very simple; yet it is responsible for a large array of high-level phenomena with enormous apparent complexity.  Of particular importance is the emergence of the proton mass-scale, $m_p \approx 1\,$GeV, which is two orders-of-magnitude larger than that associated with the Higgs mechanism of mass generation in the light-quark sector: empirically, the scale of the Higgs effect for light quarks is only $\sim 1\,$MeV.   This also has implications for the pion.  Absent a Higgs mechanism, the pion is massless, $m_\pi = 0$; but the current-masses of the light quarks in the pion are the same as they are in nucleons.  Hence, the na\"{\i}ve Higgs-mechanism result is $m_\pi \approx (m_u + m_d)$, yielding a value which is just 5\% of the physical mass.

The physical pion mass is achieved differently, being obtained via an enhancement factor, produced by dynamical chiral symmetry breaking (DCSB), which multiplies the current-quark mass contribution to the pion mass-squared \cite{GellMann:1968rz}:
\begin{equation}
m_\pi^2 = (m_u + m_d) \frac{-\langle \bar q q \rangle}{f_\pi^2}\,,
\end{equation}
where $\langle \bar q q \rangle$ is the chiral condensate \cite{Brodsky:2012ku} and $f_\pi$ is the pion's leptonic decay constant, both of which are order parameters for DCSB.

The scale of DCSB is $M_\chi \sim m_p/3$, \emph{i.e}.\ the size of a typical constituent-mass for a $u$- or $d$-quark; and the Nambu-Goldstone-boson character of the pion \cite{Nambu:1960tm, Goldstone:1961eq} means that although it \emph{should} have a mass similar to that of the $\rho$-meson, $m_\rho \approx 2 M_\chi$,  most of that mass is cancelled by gluon binding effects \cite{Roberts:2016vyn}.

In quantum mechanics the $\rho$-meson may be viewed as the valence-quark spin-flip partner of the pion.  Hence, marked differences between the properties of these two states, such as that between their masses, or unexpected similarities, which calculations might reveal, could point to features of Nature that depend critically on the properties of strong-coupling yet asymptotically free quantum field theories in four spacetime dimensions; in particular, how mass emerges.

Electromagnetic form factors should also shed light on the environment sensitivity of phenomena deriving from the emergence of mass.  For the pion, despite the experimental challenges, elastic \cite{Volmer:2000ek, Horn:2006tm, Horn:2007ug, Huber:2008id, Blok:2008jy} and transition \cite{Behrend:1990sr, Gronberg:1997fj, Aubert:2009mc, Uehara:2012ag} form factors have been measured; and the theoretical discussion of these data continues to provide novel insights \cite{Mikhailov:2016klg, Raya:2015gva, Raya:2016yuj, Horn:2016rip, Chen:2018rwz, Ding:2018xwy} and plans for new measurements \cite{Horn:2017csb, Kou:2018nap, Aguilar:2019teb}.

The short lifetime of the $\rho$-meson means that related measurements are generally impractical, although there is an empirically-based estimate of the associated magnetic moment \cite{Gudino:2015kra}: $\mu_\rho = 2.15 \pm 0.5$.  Notwithstanding the absence of experimental data, there are many theoretical computations of $\rho$-meson electromagnetic form factors, using a wide variety of tools \cite{deMelo:2018bfo}.

In addition to the interest in developing theoretical insights by contrasting $\pi$- and $\rho$-meson properties, the $J=1$ character of the $\rho$ entails that it has three distinct electromagnetic form factors and hence more structural freedom.  For instance, the $\rho$ has a quadrupole form factor; thus, like the deuteron, it possesses an observable (in principle) spectroscopic deformation \cite{Alexandrou:2012da}.  Moreover, its electric form factor, $G_E^\rho$, is the sum of three terms, one of which is negative-definite;  hence $G_E^\rho$ may possess a zero.  This possibility establishes its role as a proxy for the proton's electric form factor, for which data obtained at the Thomas Jefferson National Accelerator Facility (JLab) show a trend toward zero with increasing momentum-transfer-squared \cite{Jones:1999rz, Gayou:2001qd, Punjabi:2005wq, Puckett:2010ac, Puckett:2011xg}.

These observations provide ample motivation for the study of vector meson form factors.  Herein, therefore, we employ a continuum approach to quark-antiquark bound-states in quantum field theory, used successfully to predict and explain a wide range of hadron properties, \emph{e.g}.\, \mbox{Refs.}\,\cite{Raya:2015gva, Raya:2016yuj, Horn:2016rip, Chen:2016bpj, Eichmann:2016yit, Chen:2018rwz, Gao:2017mmp, Ding:2018xwy, Ding:2019lwe, Qin:2018dqp, Wang:2018kto, Binosi:2018rht, Qin:2019hgk, EICHMANN2019134855, Wallbott:2019dng}, to calculate the elastic electromagnetic form factors of the $\rho$-meson and study their evolution with current-quark mass.
We also compute $K^\ast$ elastic form factors; and, where worthwhile, make comparisons with the charge distributions within pseudoscalar mesons.
Our approach to the calculation of meson form factors is detailed in Sec.\,\ref{SecVMFFs}.  It produces the results discussed in Sec.\,\ref{SecVMFFsResults}.  A summary and perspective is presented in Sec.\,\ref{epilogue}.

\section{Vector Meson Form Factors: Elements}
\label{SecVMFFs}
\subsection{Form Factor Definitions}
A $J^{P}=1^{-}$ vector meson, ${\mathpzc V}$, with mass $m_{\mathpzc V}$, constituted from a valence-quark with flavour $f$ and valence-antiquark with flavour $\bar g$, has three elastic form factors and we follow Refs.\,\cite{Arnold:1979cg, Bhagwat:2006pu, Roberts:2011wy, Eichmann:2011ec} in defining them.  Denoting the incoming photon momentum by $Q$, and the incoming and outgoing ${\mathpzc V}$-meson momenta by $p^i= K-Q/2$ and $p^f= K+Q/2$, then $K\cdot Q=0$, $K^2+Q^2/4= -m_{\mathpzc V}^2$ and the ${\mathpzc V}$-$\gamma$-${\mathpzc V}$ vertex can be expressed:
\begin{subequations}
\label{Lambdarho}
\begin{align}
\Lambda_{\lambda,\mu\nu}(K,Q) & =  \sum_{q=f,g}\,e_q\,\Lambda_{\lambda,\mu\nu}^q(K,Q)\,,\\
\Lambda_{\lambda,\mu\nu}^q(K,Q) & =  \sum_{j=1}^3 T_{\lambda,\mu\nu}^{j}(K,Q) \,F_j^q(Q^2)\,,\\
F_j^{\mathpzc V}(Q^2) & =   \sum_{q=f,g}\,F^q_j(Q^2)\,,
\end{align}
\end{subequations}
where $\{e_q,q=f,\bar g\}$ are the electric charges of the valence constituents, defined in units of the positron charge, and the basis tensors are
{\allowdisplaybreaks
\begin{subequations}
\begin{align}
T_{\lambda,\mu\nu}^1(K,Q) & =  2 K_\lambda\, {\cal P}^T_{\mu\alpha}(p^i) \, {\cal P}^T_{\alpha\nu}(p^f)\,,\\
T_{\lambda,\mu\nu}^2(K,Q) & =  \left[Q_\mu - p^i_\mu \frac{Q^2}{2 m_{\mathpzc V}^2}\right] {\cal P}^T_{\lambda\nu}(p^f) \nonumber \\
& - \left[Q_\nu + p^f_\nu \frac{Q^2}{2 m_{\mathpzc V}^2}\right] {\cal P}^T_{\lambda\mu}(p^i)\,, \\
\nonumber
T_{\lambda,\mu\nu}^3(K,Q) & = \frac{K_\lambda}{m_{\mathpzc V}^2}\, \left[Q_\mu - p^i_\mu \frac{Q^2}{2 m_{\mathpzc V}^2}\right] \left[Q_\nu + p^f_\nu \frac{Q^2}{2 m_{\mathpzc V}^2}\right] \,,\\
\end{align}
\end{subequations}}
\hspace*{-0.5\parindent}where ${\cal P}^T_{\mu\nu}(p) = \delta_{\mu\nu} - p_\mu p_\nu/p^2$.   So long as a symmetry-preserving regularisation and renormalisation scheme is implemented at every stage of the calculation, the following identities are preserved:
\begin{eqnarray}
Q_\lambda \Lambda^q_{\lambda,\mu\nu}(K,Q) &=& 0\,,\\
p^i_\mu \Lambda^q_{\lambda,\mu\nu}(K,Q) &=& 0 = p^f_\nu \Lambda^q_{\lambda,\mu\nu}(K,Q)\,.
\end{eqnarray}

The electric, magnetic and quadrupole form factors are constructed as follows:
\begin{subequations}
\begin{align}
G_E^{\mathpzc V}(Q^2) & =  F_1^{\mathpzc V}(Q^2)+\frac{2}{3} \eta G_Q^{\mathpzc V}(Q^2)\,,\\
G_M^{\mathpzc V}(Q^2) & =  - F_2^{\mathpzc V}(Q^2)\,, \label{Definemu}\\
%
G_Q^{\mathpzc V}(Q^2) & =  F_1^{\mathpzc V}(Q^2) + F_2^{\mathpzc V}(Q^2) + \left[1+\eta\right] F_3^{\mathpzc V}(Q^2)\,,
\end{align}
\end{subequations}
where $\eta=Q^2/[4 m_{\mathpzc V}^2]$.  In the limit $Q^2\to 0$, these form factors define the charge, and magnetic and quadrupole moments of the ${\mathpzc V}$-meson; viz.,
\begin{subequations}
\begin{eqnarray}
\label{chargenorm}
G_E^{\mathpzc V}(Q^2=0) & = & 1\,, \\
G_M^{\mathpzc V}(Q^2=0) & = & \mu_{\mathpzc V},\;
G_Q^{\mathpzc V}(Q^2=0) = Q_{\mathpzc V}\,.
\end{eqnarray}
\end{subequations}
Furthermore, $G_E(Q^2=0)=F_1(Q^2=0)$ and
\begin{equation}
\label{Vertex0}
\Lambda(K,Q) \stackrel{Q^2\to 0}{=} 2 K_\lambda\, {\cal P}^T_{\mu\alpha}(K) \, {\cal P}^T_{\alpha\nu}(K)\, F_1(0)\,.
\end{equation}
Naturally, $F_1(0) \equiv 1$ for a meson with unit positive electric charge and $F_1(0) \equiv 0$ for a neutral meson.

It remains to specify the photon-quark interaction vertices, $\{\Lambda_{\lambda,\mu\nu}^q, q=f,\bar g\}$; and at leading order in the systematic, symmetry-preserving Dyson-Schwinger equation (DSE) approximation scheme introduced in \mbox{Refs}.\,\cite{Munczek:1994zz, Bender:1996bb}, \emph{viz}.\ rainbow-ladder (RL) truncation:
\begin{subequations}
\label{RLFV}
\begin{align}
&\Lambda_{\lambda,\mu\nu}^f(K,Q)  =
N_c {\rm tr}_{\rm D} \! \int\frac{d^4 k}{(2\pi)^4} \,
i\bar\Gamma_\nu(k;-p_f) \, S_f(k_{++}) \nonumber \\
& \times
\, i\Gamma^{f}_\lambda(k_{++},k_{-+}) \, S_f(k_{-+}) \,
i\Gamma_\mu(k_{-0};p_i) \,
S_g(k_{--})\,,\\
&\Lambda_{\lambda,\mu\nu}^g(K,Q)  =
N_c {\rm tr}_{\rm D} \! \int\frac{d^4 k}{(2\pi)^4} \,
i\bar\Gamma_\nu(k;-p_f) \, S_f(k_{++}) \nonumber \\
& \times
 i\Gamma_\mu(k_{+0};p_i) \,
S_g(k_{+-})\, i\Gamma_\lambda^g(k_{+-},k_{--})\, S_g(k_{--})\,,
\end{align}
\end{subequations}
where: $k_{\alpha \beta} = k + \alpha Q /2 + \beta p^i /2$.
The other elements in Eq.\,\eqref{RLFV} are the dressed-quark propagators, $S_{f,g}$, which, consistent with Eq.\,\eqref{RLFV}, are computed using the rainbow-truncation gap equation;
and the vector-meson Bethe-Salpeter amplitude $\Gamma_\mu(k;P)$ and amputated dressed-quark-photon vertices, $\Gamma_\lambda^{f,g}(k_f,k_i)$, both computed in RL truncation.\footnote{The impact of corrections to the RL computation is understood \cite{Raya:2015gva, Raya:2016yuj}.  The dominant effect is a modification of form factor anomalous dimensions and hence the associated logarithmic running. That running is slow and immaterial to the present discussion; but its effect can readily be incorporated when important.}

\subsection{Interaction Kernel}
\label{Interaction}
The leading-order DSE result for the vector meson form factors is now determined once an interaction kernel is specified for the RL Bethe-Salpeter equation.  We use that explained in Ref.\,\cite{Qin:2011dd, Qin:2011xq}:
\begin{subequations}
\label{KDinteraction}
\begin{align}
\mathscr{K}_{\alpha_1\alpha_1',\alpha_2\alpha_2'} & = {\mathpzc G}_{\mu\nu}(k) [i\gamma_\mu]_{\alpha_1\alpha_1'} [i\gamma_\nu]_{\alpha_2\alpha_2'}\,,\\
 {\mathpzc G}_{\mu\nu}(k) & = \tilde{\mathpzc G}(k^2) {\cal P}^T_{\mu\nu}(k) \,,
\end{align}
\end{subequations}
with ($s=k^2$)
\begin{align}
\label{defcalG}
 \tfrac{1}{Z_2^2}\tilde{\mathpzc G}(s) & =
 \frac{8\pi^2}{\omega^4} D e^{-s/\omega^2} + \frac{8\pi^2 \gamma_m \mathcal{F}(s)}{\ln\big[ \tau+(1+s/\Lambda_{\rm QCD}^2)^2 \big]}\,,
\end{align}
where $\gamma_m=12/25$, $\Lambda_{\rm QCD}=0.234\,$GeV, $\tau={\rm e}^2-1$, and ${\cal F}(s) = \{1 - \exp(-s/[4 m_t^2])\}/s$, $m_t=0.5\,$GeV.  $Z_2$ is the dressed-quark wave function renormalisation constant.
We employ a mass-independent momentum-subtraction renormalisation scheme for the gap and inhomogeneous vertex equations, implemented by making use of the scalar Ward-Green-Takahashi identity and fixing all renormalisation constants in the chiral limit \cite{Chang:2008ec}, with renormalisation scale $\zeta=2\,$GeV$=:\zeta_2$.  

The development of Eqs.\,\eqref{KDinteraction}, \eqref{defcalG} is summarised in Ref.\,\cite{Qin:2011dd} and their connection with QCD is described in Ref.\,\cite{Binosi:2014aea}; but it is worth reiterating some points.  For instance, the interaction is deliberately consistent with that determined in studies of QCD's gauge sector, which indicate that the gluon propagator is a bounded, regular function of spacelike momenta that achieves its maximum value on this domain at $s=0$ \cite{Bowman:2004jm, Boucaud:2011ug, Aguilar:2012rz, Binosi:2014aea, Binosi:2016xxu, Binosi:2016nme, Gao:2017uox, Rodriguez-Quintero:2018wma}, and the dressed-quark-gluon vertex does not possess any structure which can qualitatively alter these features \cite{Skullerud:2003qu, Bhagwat:2004kj, Aguilar:2014lha, Williams:2015cvx, Binosi:2016rxz, Binosi:2016wcx, Aguilar:2016lbe, Bermudez:2017bpx, Cyrol:2017ewj}.
It is specified in Landau gauge because, \emph{e.g}.\ this gauge is a fixed point of the renormalisation group and ensures that sensitivity to differences between \emph{Ans\"atze} for the gluon-quark vertex are least noticeable, thus providing the \mbox{conditions} for which rainbow-ladder truncation is most accurate.  
The interaction also preserves the one-loop renormalisation group behaviour of QCD so that, \emph{e.g}.\ the quark mass-functions produced are independent of the renormalisation point.
On the other hand, in the infrared, \emph{i.e}.\ $s \lesssim m_p^2$, Eq.\,\eqref{defcalG} defines a two-parameter model, the details of which determine whether confinement and/or DCSB are realised in solutions of the quark gap equations.

Computations \cite{Qin:2011dd, Qin:2011xq, Ding:2018xwy} reveal that many properties of light-quark ground-state vector- and pseudoscalar-mesons are practically insensitive to variations of $\omega \in [0.4,0.6]\,$GeV, so long as
\begin{equation}
 \varsigma^3 := D\omega = {\rm constant}.
\label{Dwconstant}
\end{equation}
This feature also extends to numerous characteristics of the nucleon and $\Delta$-baryon \cite{Eichmann:2008ef, Eichmann:2012zz, Qin:2018dqp, Wang:2018kto, Qin:2019hgk}.  In the light quark sector, therefore, the value of $\varsigma$ is chosen to reproduce, as well as possible, the measured value of the pion's mass and leptonic decay constant.  In RL truncation this requires \cite{Chen:2018rwz}
\begin{equation}
\label{varsigmalight}
\varsigma_q  =0.82\,{\rm GeV}\,,
\end{equation}
with renormalisation-group-invariant current-quark mass
\begin{equation}
\label{upmass}
\hat m_u = \hat m_d = \hat m = 6.8\,{\rm MeV}\,,
\end{equation}
which corresponds to a one-loop evolved mass of $m^{\zeta_2} = 4.7\,$MeV.  
Thus defined, one obtains $m_\pi = 0.14\,$GeV, $f_\pi = 0.095\,$GeV.

The same value of $\varsigma$ also serves for systems involving $s$-quarks.  For instance, with
\begin{equation}
\label{smass}
\hat m_s = 0.16\,{\rm GeV}\,,
\end{equation}
corresponding to $m_s^{\zeta_2} = 0.11\,$GeV, $\varsigma_q$ in Eq.\,\eqref{varsigmalight} produces
a good description of $K$, $\eta$, $\eta^\prime$ physics \cite{Chen:2018rwz, Ding:2018xwy}.

Herein, we also consider properties of vector mesons constituted from a degenerate quark and antiquark whose mass matches that of the $c$-quark.  It is therefore pertinent to remark that RL truncation has been explored in connection with heavy-light mesons and heavy-quarkonia \cite{Bhagwat:2004hn, Hilger:2014nma, Ding:2015rkn, Gomez-Rocha:2016cji, Chen:2016bpj, Hilger:2017jti, Binosi:2018rht}.  Those studies reveal that improvements to RL can be important in heavy-light systems; and a RL-kernel interaction strength fitted to pion properties alone is not optimal in the treatment of heavy quarkonia.  Both observations are readily understood, but we focus on the latter because it is most relevant to this study.

Recall, therefore, that for meson bound-states it is now possible \cite{Chang:2009zb, Chang:2010hb, Chang:2011ei} to employ sophisticated kernels which overcome many weaknesses of RL truncation.  The new technique is symmetry preserving and has an additional strength, \emph{i.e}.\ the capacity to express DCSB nonperturbatively in the integral equations connected with bound-states.  Owing to this feature, the scheme is described as the ``DCSB-improved'' or ``DB'' truncation.  In a realistic DB truncation, $\varsigma^{\rm DB} \approx 0.6\,$GeV; a value which coincides with that predicted by solutions of QCD's gauge-sector gap equations \cite{Binosi:2014aea, Binosi:2016wcx, Binosi:2016nme, Rodriguez-Quintero:2018wma}.
Straightforward analysis shows that corrections to RL truncation largely vanish in the heavy+heavy-quark limit;
%
%
hence the aforementioned agreement entails that RL truncation should provide a sound approximation for systems involving only heavy-quarks so long as one employs $\varsigma^{\rm DB}$ as the infrared mass-scale.  In heavy-quark systems we therefore employ \mbox{Eqs}.\,\eqref{KDinteraction}, \eqref{defcalG} as obtained using
\begin{equation}
\label{varsigmaQ}
\varsigma_Q = 0.6\,{\rm GeV}\,,
\end{equation}
with $\omega=0.8\,$GeV \cite{Chen:2016bpj, Qin:2018dqp, Binosi:2018rht}.
In this case, a renormalisation-group-invariant current-quark mass
\begin{equation}
\label{cmass}
\hat m_{c}=1.75\,{\rm GeV}\,,
\end{equation}
corresponding to the one-loop-evolved value
$m_{c}^{2\,{\rm GeV}}=1.21\,{\rm GeV}$,
yields $m_{J/\Psi} = 3.09\,$GeV, $f_{J/\Psi}=0.29\,$GeV, values which compare favourably with other determinations, respectively: $3.10\,$GeV \cite{Tanabashi:2018oca} and $0.286(4)$ \cite{Donald:2012ga}.

\subsection{Propagators, Amplitudes and Vertices}
The RL approximation to the elastic electromagnetic form factor of a vector meson with mass $m_{\mathpzc V}$ is now obtained as follows.
(\emph{i}) Solve the dressed-quark gap equation using the interaction and current-quark masses specified in Sec.\,\ref{Interaction}, following Ref.\,\cite{Maris:1997tm} and adapting the algorithm improvements from Ref.\,\cite{Krassnigg:2009gd} when necessary.
(\emph{ii}) With the dressed-quark propagators obtained thereby and the same interaction, solve the inhomogeneous Bethe-Salpeter equations to obtain the unamputated dressed-quark-photon vertices, including their dependence on $Q^2$, as described, \emph{e.g}.\ in Ref.\,\cite{Maris:1999bh}.
(\emph{iii}) With the same inputs, solve the homogeneous Bethe-Salpeter equations to obtain the (amputated) Bethe-Salpeter amplitudes for each of the desired vector meson bound-states, obtaining a complete picture of their dependence on $(k^2,k\cdot P)$.
(\emph{iv}) Combine these elements to form the integrands in Eq.\,\eqref{RLFV} and compute the integrals as a function of $Q^2$ to extract the form factors, $F_{1,2,3}^{\mathpzc V}(Q^2)$.
%

It is here worth recording the following remarks.
$(\mathpzc a)$--A vector meson Bethe-Salpeter amplitude involves eight independent scalar functions, each labelled by the bound-state mass-squared and depending on $(k^2,k\cdot P)$ \cite{LlewellynSmith:1969az}.  The Bethe-Salpeter wave function, constructed by attaching the external dressed-quark propagator legs to the amplitude, is expressed in terms of eight analogous functions.  In the meson's rest frame, four of these functions describe $^3S_1$ orbital angular momentum correlations between the dressed valence quarks and the other four describe $^3D_1$ correlations.  Typical solutions of the vector meson bound-state problem indicate that the $^3D_1$-wave strength is large \cite{Maris:1999nt, Krassnigg:2009zh, Gao:2014bca, Hilger:2014nma}; hence, vector mesons are deformed.
$(\mathpzc b)$--In completing steps (\emph{iii}) and (\emph{iv}), we emulate Ref.\,\cite{Bhagwat:2006pu} and solve directly for both $\Gamma_\mu(k_{+0},p_i=K-Q/2)$ and $\Gamma_\nu(k;p_f=K+Q/2)$ at each value of $Q^2$ for which the form factor is desired.  This procedure is time consuming but it improves numerical accuracy at higher $Q^2$-values.
$(\mathpzc c)$--In a properly implemented RL truncation, \emph{i.e}.\ so long as one employs a symmetry-preserving regularisation scheme in solving for the propagators, amplitudes and vertices, then, for a unit-charge state, $F_1^{\mathpzc V}(0)=1=G_E^{\mathpzc V}(0)$ is guaranteed by the canonical normalisation of the vector meson Bethe-Salpeter amplitude and Eq.\,\eqref{Vertex0} reproduces the standard photo-interaction vertex for an on-shell vector meson.

\section{Vector Meson Form Factors: Results}
\label{SecVMFFsResults}
\subsection{Static Properties}
We have computed the elastic electromagnetic form factors of the charged $\rho$-meson, charged and neutral $K^\ast$-meson, and those of fictitious charged vector-mesons constituted from $u$- and $\bar d$-like quarks with current masses equal to those of the $s$- and $c$-quarks, \emph{viz}.\ $\rho_s$ and $\rho_c$, respectively.  Results for the static properties of these systems are collected in Table~\ref{static}.

\begin{table}[t]
\caption{\label{static}
Calculated values for a range of vector-meson static properties: quadrature integration error is $\lesssim 2$\%. %
The $\rho$-meson is built from mass-degenerate valence-quarks with the current-masses in Eq.\,\eqref{upmass}; $\rho_s$, using Eq.\,\eqref{smass}; and $\rho_c$, using Eq.\,\eqref{cmass}.
The radii are defined in \mbox{Eq}.\,\eqref{eqradii}.
For comparison:
an average of computed $\rho$-meson results tabulated elsewhere \cite{deMelo:2018bfo} yields
$r_\rho = 0.67(12)\,$fm, $\mu_\rho=2.17(21)$, $Q_\rho = -0.55(28)$;
and using an interaction similar to that defined by Eqs.\,\eqref{KDinteraction}, \eqref{defcalG}, Ref.\,\cite{Bhagwat:2006pu} reports
$r_\rho = 0.73\,$fm, $\mu_\rho=2.01$, $Q_\rho = -0.41$,
$r_{\rho_c} = 0.23\,$fm, $\mu_{\rho_c}=2.13$, $Q_{\rho_c} = -0.28$.
A lattice-QCD (lQCD) simulation yields \cite{Owen:2015gva} $r_\rho = 0.82(4)\,$fm,  $\mu_\rho=2.21(8)$;
and another produces \cite{Dudek:2006ej} $r_{\rho_c} = 0.257(4)\,$fm, $\mu_{\rho_c} = 2.10(3)\,$fm,  $Q_{\rho_c}=-0.23(2)$.
For pointlike vector mesons with unit charge \cite{Brodsky:1992px, Haberzettl:2019qpa}: $\mu = 2$, $Q = -1$.
%
%
Where known, empirical values are \cite{Tanabashi:2018oca}:
$m_\rho = 0.775\,$GeV, $f_\rho = 0.156(1)\,$GeV,
$m_{K^\ast} = 0.892\,$GeV, $f_{K^\ast} = 0.158(8),$GeV,
$m_\phi = 1.019\,$GeV, $f_\phi = 0.161(3)\,$GeV,
$m_{J/\Psi} = 3.097\,$GeV; and
a lQCD study obtains $f_{J/\Psi} = 0.286(4)\,$GeV \cite{Donald:2012ga}.
(Insofar as masses and leptonic decay constants are concerned, $\rho_{s,c}$ results can be compared with those for $\phi$- and $J/\Psi$- mesons.)
}
\begin{tabular*}
{\hsize}
{
|l@{\extracolsep{0ptplus1fil}}
|c@{\extracolsep{0ptplus1fil}}
c@{\extracolsep{0ptplus1fil}}
c@{\extracolsep{0ptplus1fil}}
c@{\extracolsep{0ptplus1fil}}
c|@{\extracolsep{0ptplus1fil}}}\hline
${\mathpzc V}\ $    & $\rho$ & $K^{+\ast}$ & $K^{0\ast}$ & $\rho_s$ & $\rho_c$ \\\hline
$m_{\mathpzc V}/{\rm GeV}\ $ & $\phantom{-}0.75\ $ & \multicolumn{2}{c}{$0.96\ $ }&  $\phantom{-}1.08\ $ & $\phantom{-}3.09\ $ \\
$f_{\mathpzc V}/{\rm GeV}\ $ & $\phantom{-}0.15\ $ & \multicolumn{2}{c}{$0.17\ $ } & $\phantom{-}0.19\ $& $\phantom{-}0.29\ $\\
$r_{\mathpzc V}/{\rm fm}\ $ & $\phantom{-}0.72\ $& $\phantom{-}0.64\ $& $\phantom{-}0.27i\ $& $\phantom{-}0.52\ $ & $\phantom{-}0.24\ $\\
$r_{\mathpzc V} m_{\mathpzc V}\ $ & $\phantom{-}2.76\ $& $\phantom{-}3.13\ $ & $\phantom{-}1.29i\ $ & $\phantom{-}2.85\ $ & $\phantom{-}3.70\ $\\
$\mu_{\mathpzc V}\ $ & $\phantom{-}2.01\ $ & $\phantom{-}2.22\ $& $-0.26\phantom{i}\ $ & $\phantom{-}2.08\ $& $\phantom{-}2.12\ $\\
$r^\mu_{\mathpzc V} m_{\mathpzc V}\ $ & $\phantom{-}2.63\ $& $\phantom{-}3.05\ $& $\phantom{-}4.40\phantom{i}\ $& $\phantom{-}2.71\ $ & $\phantom{-}3.59\ $\\
$Q_{\mathpzc V}\ $ & $-0.36\ $ & $-0.31\ $ & $-0.021\ $ & $-0.32\ $ & $-0.33\ $\\\hline
\end{tabular*}
\end{table}

The listed radii are defined as follows ($L=E,M$):
\begin{equation}
\label{eqradii}
(r_{\mathpzc V}^{L})^2 m_{\mathpzc V}^2 = -[6/G_{L}^{\mathpzc V}(0)]
\left. \frac{d}{d x} G_{L}^{\mathpzc V}(x) \right|_{x=0}.
\end{equation}
This radius-squared can be negative for neutral hadrons, which explains the ``$i$'' in such cases.  The vector-meson leptonic decay constants are computed using
\begin{align}
f_{\mathpzc V} m_{\mathpzc V} = {\rm tr}_{\rm D} Z_2\int^\Lambda \frac{d^4 k}{(2\pi)^4} \gamma_\mu S(k_{+}) \Gamma_\mu(k;P) S(k_{-})\,,
\end{align}
where $k_\pm = k\pm P/2$ and a symmetry-preserving regularisation scheme is used to define and evaluate the integral.

It is natural to first compare the vector meson charge radii with those of appropriate pseudoscalar meson analogues:
$r_\pi = 0.66\,$fm \cite{Chen:2018rwz};
$r_{K^+} = 0.56\,$fm \cite{Gao:2017mmp},
$r_{K^0} = 0.26 i \,$fm \cite{Gao:2017mmp};
$r_{\pi_s} = 0.49\,$fm \cite{Chen:2018rwz};
and
$\tilde r_{\eta_c^0} = 0.16\,$fm, which is an interaction radius defined via the $\gamma^\ast \gamma \to \eta_c$ transition form factor \cite{Raya:2016yuj}.  (Notably, $\tilde r_{\pi^0} \approx r_\pi$ \cite{Raya:2015gva}.)
Evidently, the radius of a given vector meson is larger than its pseudoscalar counterpart.  The difference diminishes with increasing current-quark mass because spin-dependent interactions are suppressed as current-quark masses grow.
(Ref.\,\cite{Bhagwat:2006pu} reports $r_{\rho_c} = 0.23\,$fm \emph{cf}.\ $r_{\pi_c} = 0.22\,$fm.)

As with pseudoscalar mesons \cite{Chen:2018rwz}, the product of decay-constant and charge-radius is roughly constant for systems composed of light quarks:
\begin{equation}
\label{ftimesr}
\hat m_{f,\bar g} \lesssim m_s \; | \;
f_{\mathpzc V} r_{\mathpzc V} = 0.53(3)\,,
\end{equation}
\emph{i.e}.\ within the domain upon which emergent mass is dominant.  Thereafter, as the Higgs-mechanism of mass generation becomes increasingly more effective: the value of $f_{\mathpzc V} r_{\mathpzc V}$ falls with increasing $\hat m$; and the analogous product for pseudoscalar mesons, $f_{\mathpzc P} r_{\mathpzc P}$,  evolves so that the two products become equal in the heavy-heavy limit \cite{Bhagwat:2006xi}.
Of course, the charge radius is defined as a measure of the behaviour of $G_E^{\mathpzc V}$ on $x\simeq 0$.  The evolution of $G_E^{\mathpzc V}$ with $x$ is discussed below.

As found elsewhere \cite{Bhagwat:2006pu}, Table~\ref{static} reveals that the dimensionless vector meson magnetic moment, Eq.\,\eqref{Definemu}, increases with increasing current-quark mass.  However, the growth is slow, becoming practically indiscernible on $\hat m_{f,\bar g} \gtrsim \hat m_s$, \emph{viz}.\ the domain upon which the Higgs mechanism of mass generation dominates.  Consequently, our predicted value remains close to $2$; hence, the interaction between an external magnetic field and the composite, nonpointlike vector meson diminishes as $\sim 2/m_{\mathpzc V}$, following the point-particle pattern.

On the other hand, the magnitude of the dimensionless quadrupole moment is significantly smaller than the point particle value and decreases slowly with increasing current-quark mass.  Accordingly, the state remains deformed and the quadrupole interaction between an external electromagnetic field and the composite, nonpointlike vector meson falls as $\sim -0.3 /m_{\mathpzc V}^2$ with growing meson mass.  Like the earlier DSE prediction \cite{Bhagwat:2006pu}, our result has the same sign as that obtained in a recent lQCD simulation and is similar in magnitude \cite{Owen:2015gva}.

In order to expose deviations from point-particle behaviour and hence deformation in the composite vector meson systems, one can write \cite{Alexandrou:2012da}:
\begin{align}
\mu_{\mathpzc V} &= \phantom{-} 2 + (\kappa_\gamma^{\mathpzc V}-1) + \lambda_\gamma^{\mathpzc V}\,, \\
Q_{\mathpzc V} &= -1 + (1-\kappa_\gamma^{\mathpzc V}) + \lambda_\gamma^{\mathpzc V} \,,
\end{align}
in terms of which point-particle behaviour is indicated by $1-\kappa_\gamma^{\mathpzc V}=0=\lambda_\gamma^{\mathpzc V}$.  Expressing the Table~\ref{static} results in this way, one finds
\begin{equation}
\label{deformation}
\begin{array}{c|c|c|c|c}
  {\mathpzc V}      & \rho & \rho_s & \rho_c & K^{\ast +} \\\hline
1-\kappa_\gamma^{\mathpzc V} & 0.32 & 0.32 & 0.28 & 0.23 \\
\lambda_\gamma^{\mathpzc V} & 0.32 & 0.37 & 0.40 & 0.46
\end{array}\,.
\end{equation}
Thus perceived, one sees roughly-equal compositeness-induced deviations/deformations in both the magnetic and quadrupole moments for all systems.

It is worth remarking here that RL truncation omits what are commonly called meson-cloud contributions (MCCs).  At realistic light-quark masses, their impact on charge radii is small and can be absorbed into the definition of $\varsigma$ \cite{Chen:2018rwz}; but MCCs may be quantitatively important for those static observables which are more sensitive to angular momentum, such as magnetic and quadrupole moments.  Importantly, MCCs diminish rapidly with increasing $Q^2$, being negligible for a typical charged-hadron electric form factor on $Q^2 \gtrsim  0.25\,$GeV$^2$ \cite{Eichmann:2011vu}.

\subsection{Focus: Electric Form Factor}
The electric form factor of a positively-charged vector meson decreases with increasing $x=Q^2/m_{\mathpzc V}^2$. However, setting it apart from that of a pseudoscalar meson, which is positive-definite, the large-$x$ prediction from Refs.\,\cite{Brodsky:1992px, Haberzettl:2019qpa} suggests that $G_E^{\mathpzc V}(x)$ may possess a zero at $x \sim 6$.  This was the outcome in Ref.\,\cite{Roberts:2011wy}, which used a symmetry preserving regularisation of a contact interaction and was thus able to compute form factors to arbitrarily large $x$.

In exhibiting a zero crossing, $G_E^{\mathpzc V}$ can serve as a surrogate for the proton's electric form factor, $G_E^{p}$, for which modern data show a trend toward zero with increasing $Q^2$ \cite{Jones:1999rz, Gayou:2001qd, Punjabi:2005wq, Puckett:2010ac, Puckett:2011xg}: linear extrapolation yields a zero in $G_E^{p}$ at
\begin{equation}
\label{protonzero}
Q_p^2 \approx 9.8\,m_p^2 = 8.7 \, {\rm GeV}^2.
\end{equation}
The reason for the potential appearance of a zero is similar in both cases.  For the proton, a zero can be produced by destructive interference between the Dirac and Pauli form factors, and will appear if the transition between the strong and perturbative domains of QCD is pushed to a sufficiently large value of $Q^2$ \cite{Wilson:2011aa, Segovia:2014aza}.  In the vector meson case, it is a destructive interference between $F_{1,3}^{\mathpzc V}$ (positive) and $F_2^{\mathpzc V}$ (negative): if the magnetic  form factor, $F_2^{\mathpzc V}$, is removed, then $G_E^{\mathpzc V}$ is positive-definite at spacelike momenta.

The merit of using vector meson studies to locate and explain a zero in the electric form factor of a $J\neq 0$ hadron is the relative simplicity of the two-body continuum bound-state problem as compared to the analogous three-body problem; but this does not make it easy.  As in most calculations of hadron form factors that have worked directly with a realistic RL quark-quark scattering kernel, we use brute-force numerical techniques.  Consequently, owing to moving singularities in the complex-$k^2$ domain sampled by the bound-state equations \cite{Maris:1997tm}, for each vector meson there is a maximum value of $Q^2$ beyond which evaluation of the integrals in Eqs.\eqref{RLFV} is no longer possible with conventional algorithms.

More sophisticated methods have been developed \cite{Chang:2013nia, Raya:2015gva, Raya:2016yuj, Gao:2017mmp, Ding:2018xwy}, based on the perturbation theory integral representation (PTIR) \cite{Nakanishi:1969ph}.  Constructing accurate PTIRs is, however, time consuming; and especially so in our case because one would need to build PTIRs for each quark propagator, Bethe-Salpeter amplitude and photon-quark vertex considered herein, \emph{i.e}.\ roughly 100 scalar functions.

We therefore persist with a straightforward RL truncation, computing all form factors on the accessible domain and then extrapolating to larger $Q^2$-values using the Schlessinger point method (SPM), whose properties and accuracy are explained elsewhere  \cite{Schlessinger:1966zz, PhysRev.167.1411, Tripolt:2016cya, Chen:2018nsg, Binosi:2018rht, Binosi:2019ecz}.  We note only that the SPM is based on the Pad\'e approximant.  It is able to accurately reconstruct a function in the complex plane within a radius of convergence specified by that one of the function's branch points which lies nearest to the real domain from which the sample points are drawn.  Moreover, owing to the procedure's discrete nature, the reconstruction can also provide a reasonable continuation on a larger domain along with an estimate of the associated error.

\begin{figure}[t]
\includegraphics[width=0.4\textwidth]{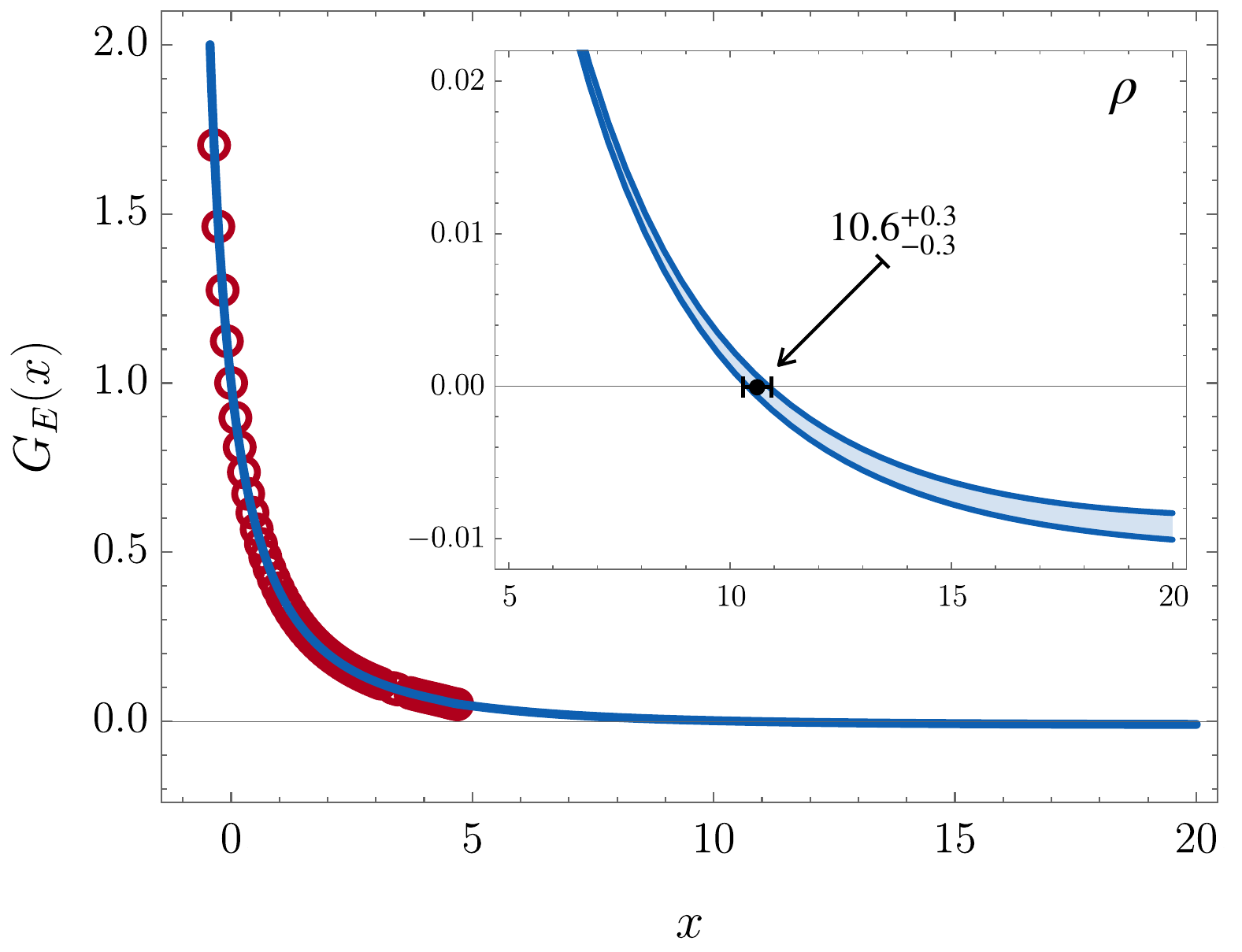}
\medskip

\includegraphics[width=0.4\textwidth]{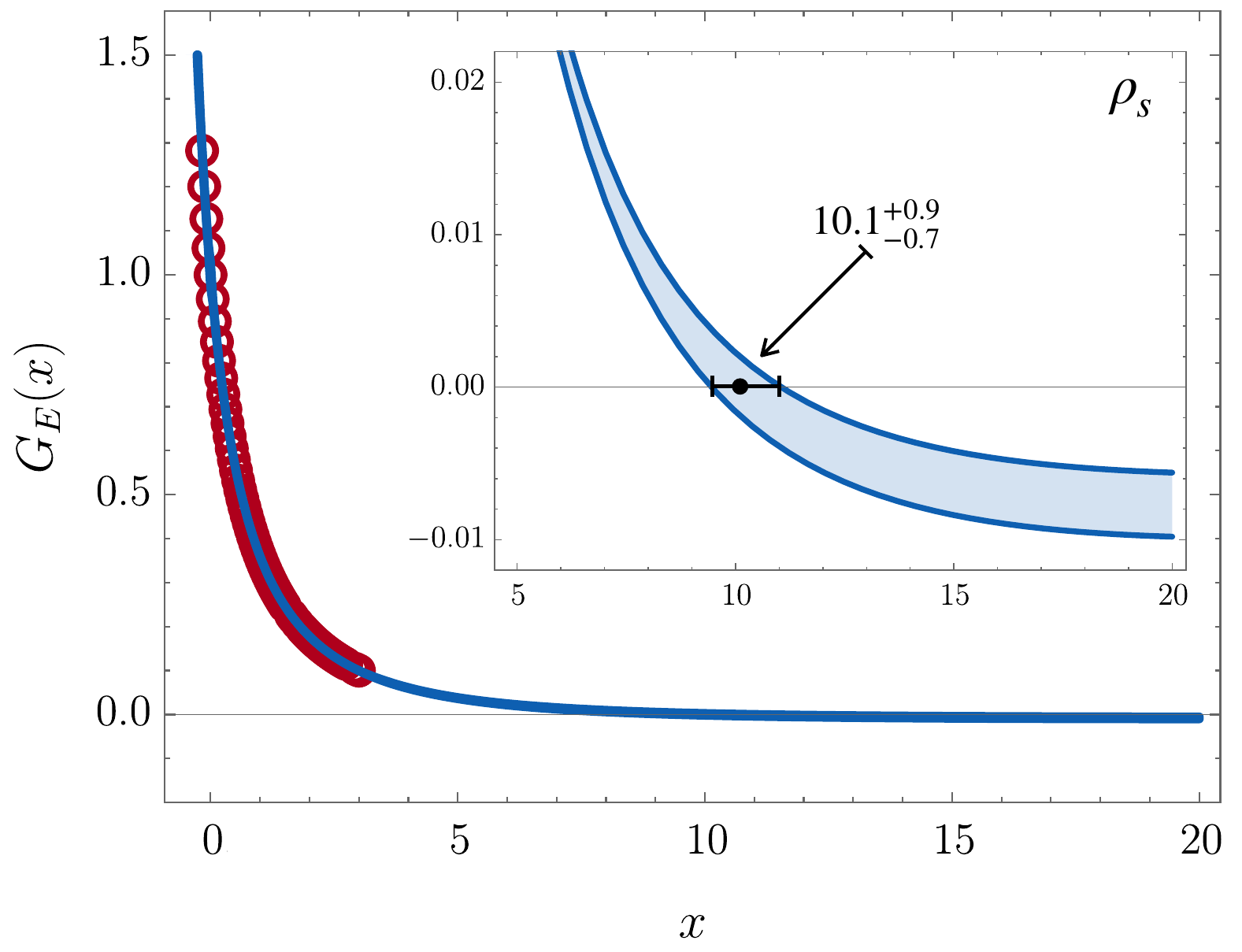}
\medskip

\includegraphics[width=0.4\textwidth]{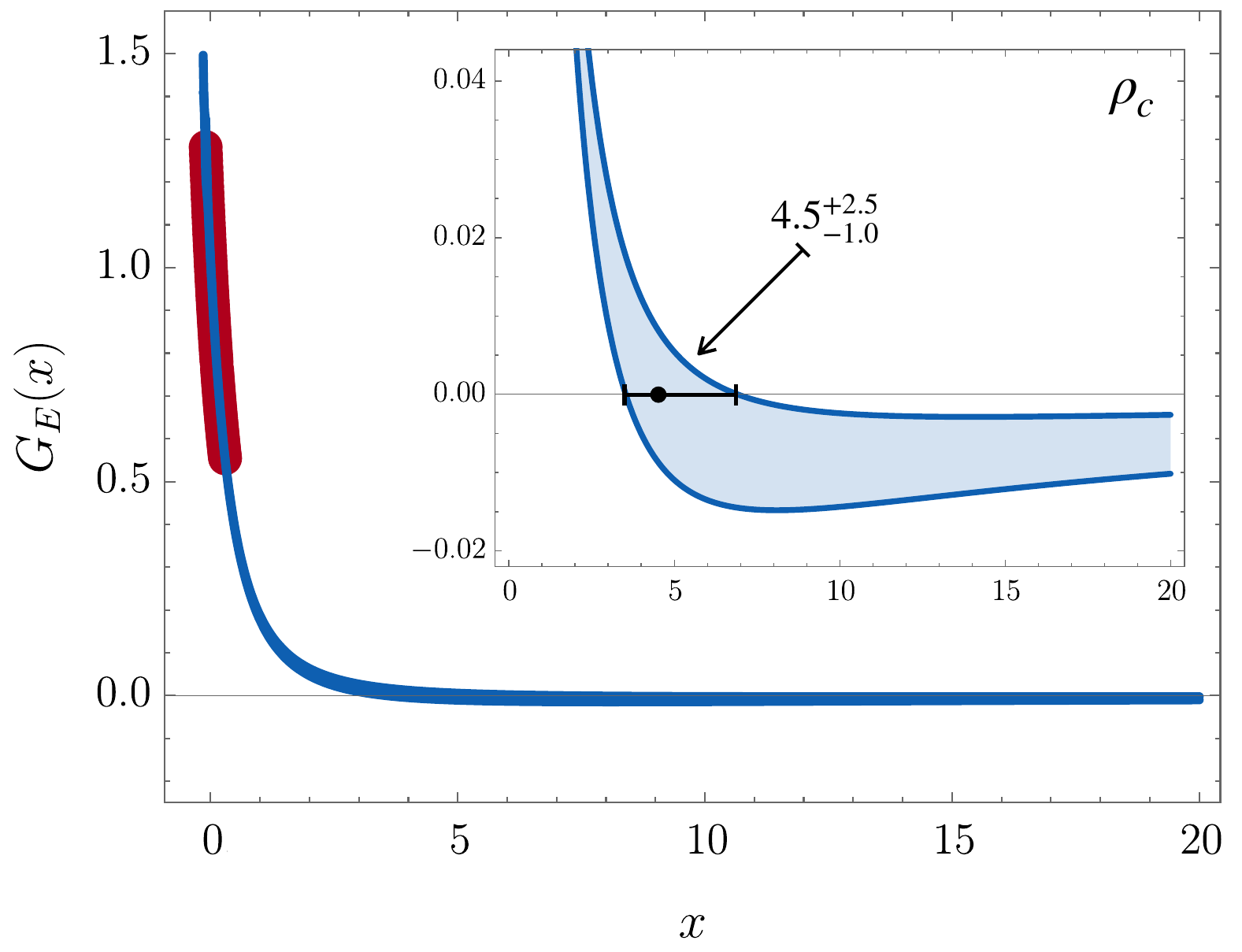}
\caption{\label{FigGErho}
Electric form factors of positively-charged vector mesons built from mass-degenerate valence-quarks: \emph{upper panel}, $\rho$ -- Eq.\,\eqref{upmass}; \emph{middle panel} $\rho_s$ -- Eq.\,\eqref{smass}; \emph{lower panel}, $\rho_c$ -- Eq.\,\eqref{cmass}.
}
\end{figure}

In the three panels of Fig.\,\ref{FigGErho}, as functions of $x=Q^2/m_{\mathpzc V}^2$, we display our computed electric form factors for the three positively-charged vector mesons in Table~\ref{static} that are built from mass-degenerate valence-quarks: ${\mathpzc V}=\rho$, $\rho_s$, $\rho_c$.  Our analysis predicts a zero in each case; and importantly, as the current-mass of the system's valence-quarks is increased, the $x$-location of the zero, $x_{\mathpzc z}$, moves toward $x=0$:
\begin{equation}
\label{GEzero}
\begin{array}{l|c|c|c}
{\mathpzc V} & \rho & \rho_s & \rho_c\\\hline
\rule{0em}{3ex} x_{\mathpzc z} & 10.6(3) & 10.1^{(9)}_{(7)}&  4.5^{(2.5)}_{(1.0)}
\end{array} .
\end{equation}
The shift is initially slow; but the pace increases as one leaves the domain upon which emergent mass is dominant and enters into that for which explicit (Higgs-connected) mass generation overwhelms effects deriving from strong-QCD dynamics.  Reverting to $Q^2$, the location of the zero in $G_E^{\mathpzc V}$ moves to larger values with increasing current-quark mass.

Interestingly, working from Eq.\,\eqref{GEzero}, replacing $m_\rho^2$ by $m_p^2$, then one is led to estimate that a zero appears in the proton's elastic form factor at $Q^2 \approx 9.4(3)\,$GeV$^2$.  This value is comparable with that in Eq.\,\eqref{protonzero} and compatible with the prediction in Ref.\,\cite{Segovia:2014aza}: $Q^2 \approx 9.5\,$GeV$^2$.

\begin{figure}[t]
\includegraphics[width=0.4\textwidth]{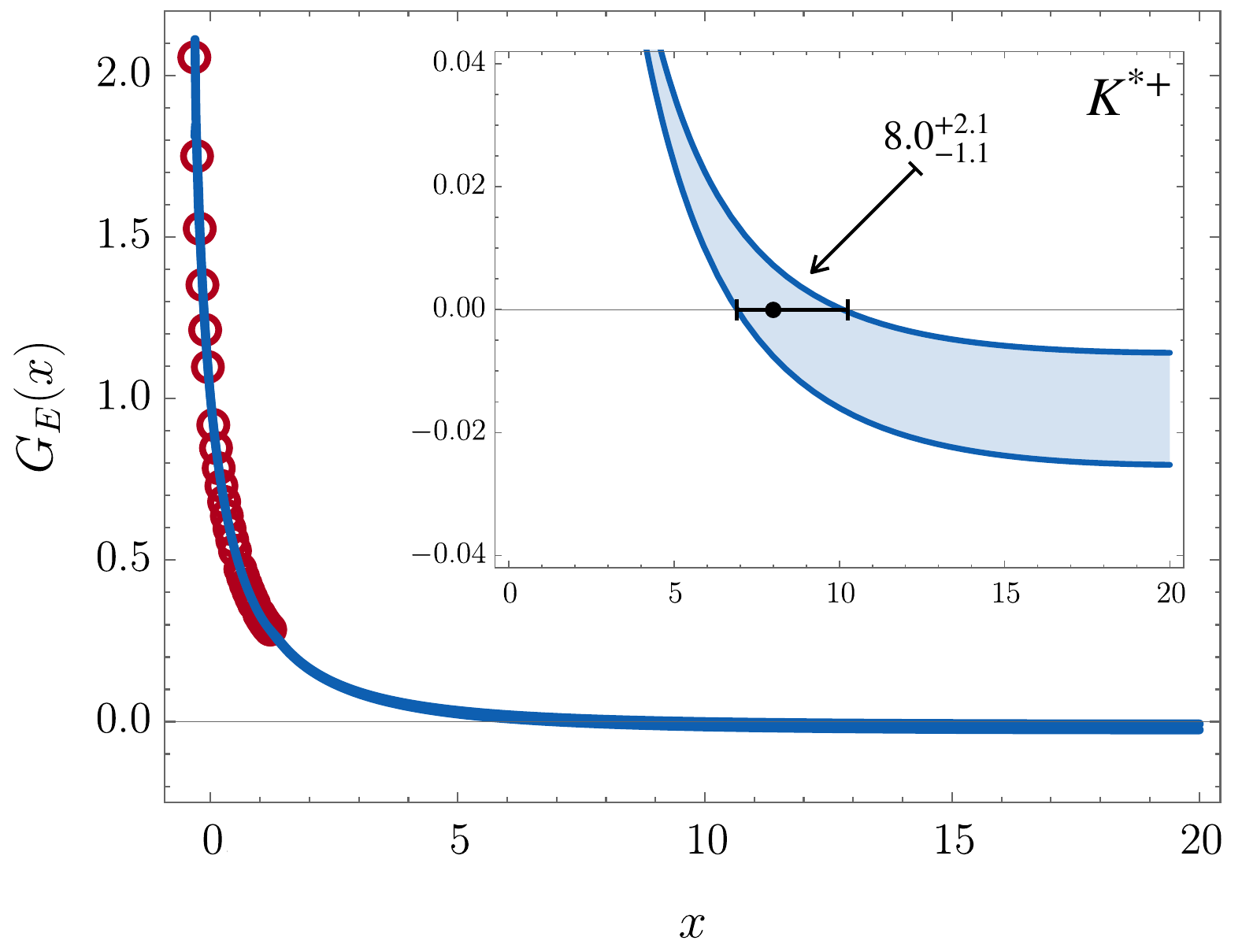}
\medskip

\includegraphics[width=0.4\textwidth]{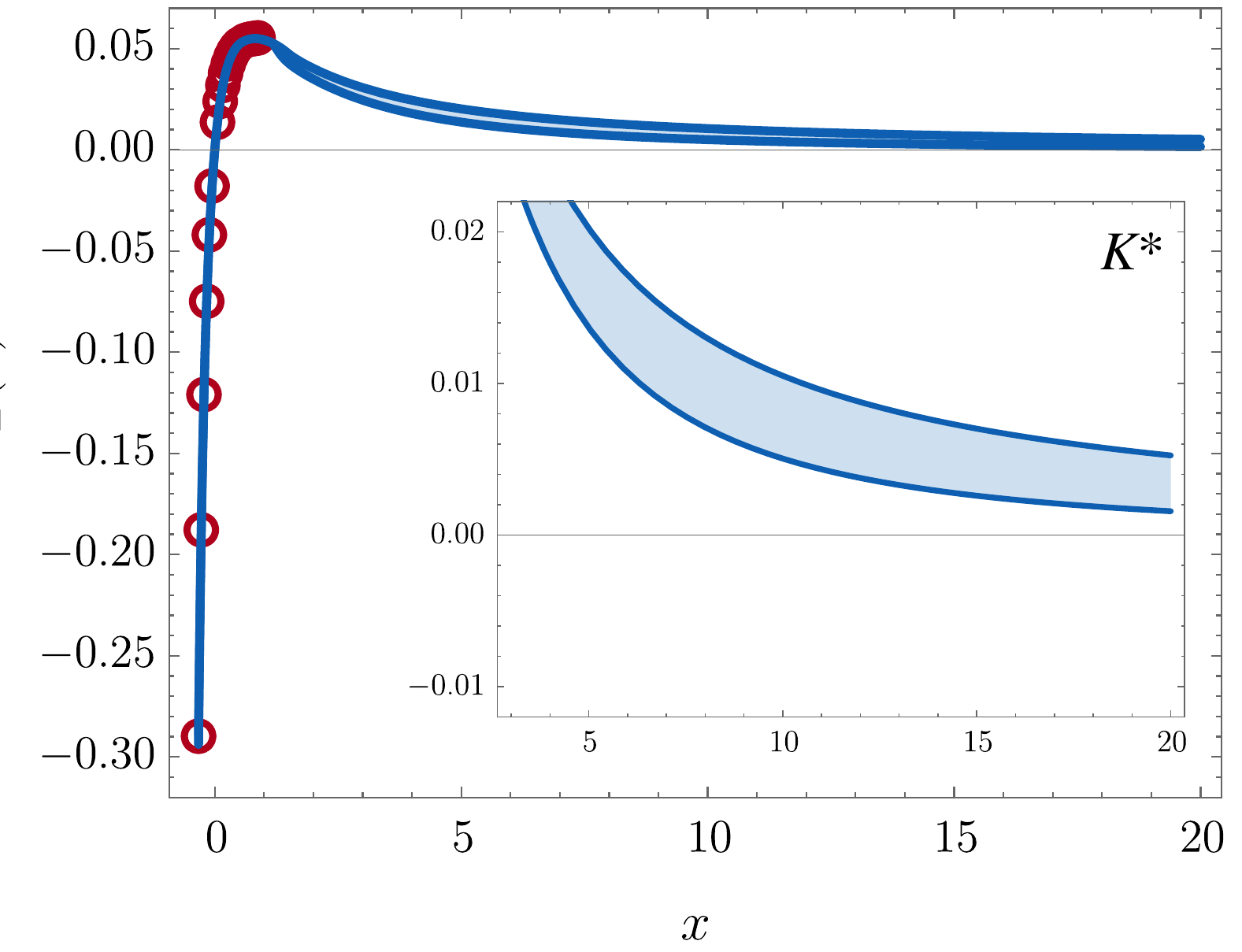}
\caption{\label{FigGEKst}
Electric form factors of $K^\ast$ mesons, with $x=Q^2/m_{K^\ast}^2$: \emph{upper panel}, electric-charge positive; and \emph{lower}, electric-charge neutral.
}
\end{figure}

In developing an understanding of these features, it is useful to bear the following observations in mind. (\emph{i}) In the cases under consideration, the RL-dressed photon-quark vertex always possesses a pole at $Q^2/m_{\mathpzc V}^2= -1$; hence, so does $G_E^{\mathpzc V}$.  (This becomes a resonance peak with the inclusion of decay channels, but that is immaterial here.)  (\emph{ii}) $G_E^{\mathpzc V}(0)=1$ for every positively-charged vector meson.  (\emph{iii}) Fig.\,\ref{FigGErho} plots $G_E^{\mathpzc V}(x=Q^2/m_{\mathpzc V}^2)$.

Recall now that the Higgs-mechanism for mass generation is dominant for heavy quarks.  Hence, in this sector the dressed-quark mass function does not run quickly; the effective quark (recoil) mass is roughly fixed at $M_c^{E} \sim m_{\mathpzc V}/2$; the scattering photon probes this scale; and the recoiling dressed-quark has a large magnetic form factor.  Consequently, $G_E^{\mathpzc V}(x)$ exhibits a zero at a given location, not too far from $x=0$.
On the other hand, DCSB drives mass generation in the light-quark sector so the dressed-quark mass runs rapidly. Hence, the recoiling system within $\mathpzc V$ has a mass which drops quickly toward zero and a magnetic form that does likewise \cite{Singh:1985sg, Bicudo:1998qb, Chang:2010hb}.  The photon probe resolves this dressed-quark, finding a recoiling target quark whose active mass and magnetic \mbox{moment} become smaller as $Q^2$ increases.  The ``effective $x$'' is therefore larger than $Q^2/m_{\mathpzc V}^2$, something expressed in an electric form factor which evolves more slowly with $x$ than might na\"{\i}vely be expected, \emph{i.e}.\ a zero located further from $x=0$.

It is also worth performing a similar analysis for the $K^\ast$ mesons; and our results are depicted in Fig.\,\ref{FigGEKst}.  In this case, owing to the imbalance between current-quark masses and consequent skewing of the ``safe domain'' of complex-plane integration for Eq.\,\eqref{RLFV}, a direct calculation of the form factors is impossible beyond $x=Q^2/m_{K^\ast}^2\approx 1.2$; hence, approximation using the SPM involves a lar\-ger error.  Notwithstanding that, one may confidently conclude that the electric form factor of the positively charged $K^\ast$ exhibits a zero at
\begin{equation}
x_{\mathpzc z}^{K^{\ast +}} =8.0^{(2.1)}_{(1.1)}\,,
\end{equation}
whereas the analogous form factor of the neutral $K^\ast$ is positive definite on $x>0$.

\begin{figure}[t]
\includegraphics[width=0.4\textwidth]{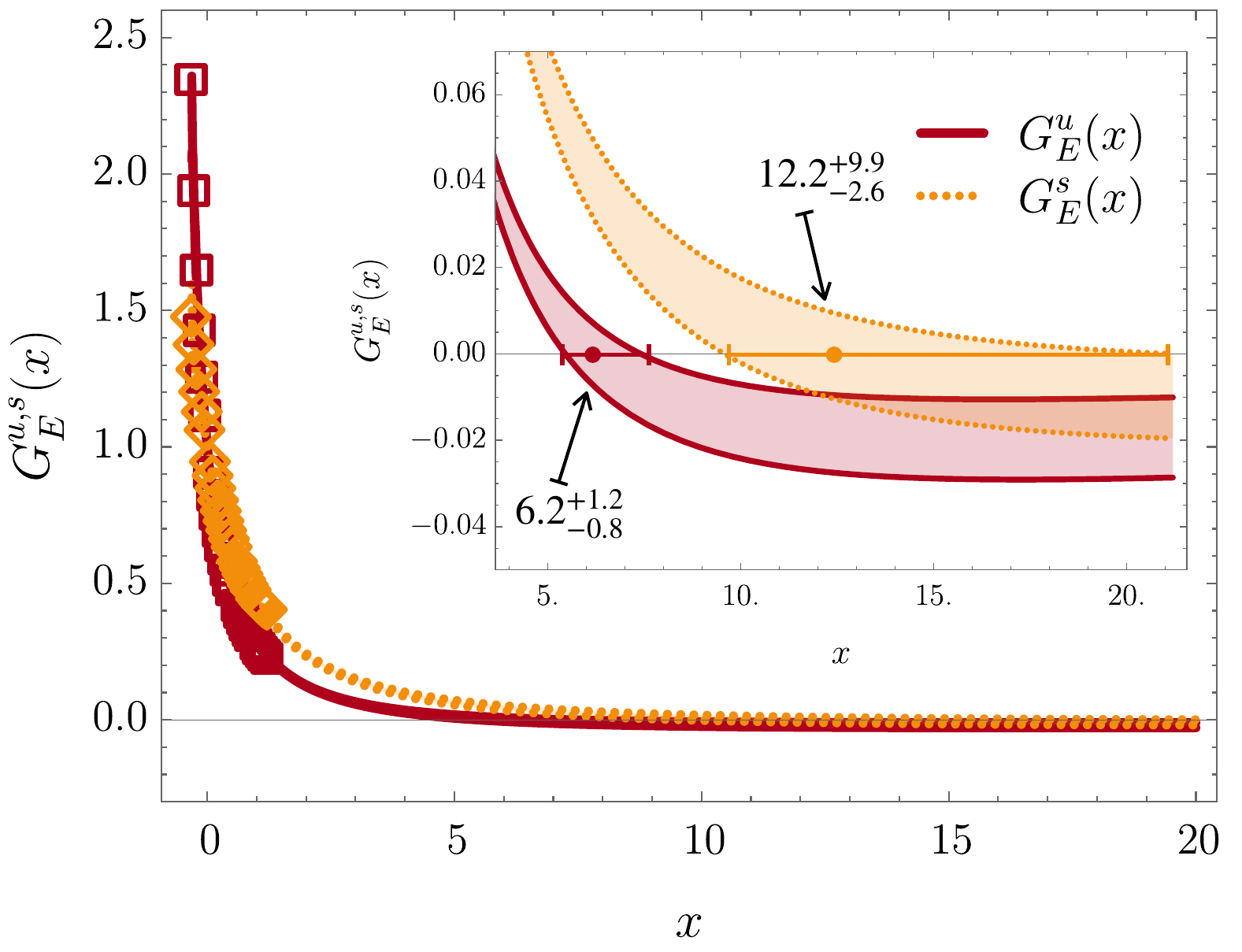}
\caption{\label{FigGEKstFlavour}
Flavour-separated $K^\ast$ electric form factors plotted versus $x=Q^2/m_{K^\ast}^2$: solid (red) curve marked by squares -- $u$-quark; dotted (orange) curve, diamonds -- $\bar s$-quark.  Electric-charge factors have been divided out so both curves are unity at $x=0$.
}
\end{figure}

Our confidence in the existence/absence of a zero in these cases is not based on Fig.\,\ref{FigGEKst}; but, instead, upon Fig.\,\ref{FigGEKstFlavour}, which displays the flavour-separated $K^\ast$ \mbox{form} factors.\footnote{Recall that we have assumed isospin symmetry.  Hence, the form factors associated with the $u$-quark in the positive-$K^\ast$ are the same as those for the $d$-quark in the neutral $K^\ast$, apart for a multiplicative factor of ``$-2$''.}
Evidently, both curves exhibit a zero:
\begin{align}
G_{Eu}^{K^\ast} \; & {\rm at} \; x=\phantom{1}6.2^{(1.2)}_{(0.8)}; \\
G_{Es}^{K^\ast} \; & {\rm at} \; x=12.2^{(9.9)}_{(2.6)}.
\end{align}
Consequently,
$G_{E}^{K^{\ast +}} = (2/3) G_{Eu}^{K^\ast} + (1/3)G_{Es}^{K^\ast}$ must also possess a zero.
On the other hand,
$G_{E}^{K^{\ast 0}} = - (1/3)G_{Eu}^{K^\ast} + (1/3)G_{Es}^{K^\ast}$ is positive-definite on $x>0$ because the zero in
$(- G_{Eu}^{K^\ast})$ occurs much before that in $G_{Es}^{K^\ast}$ and
$|G_{Eu}^{K^\ast}| > |G_{Es}^{K^\ast}|>1$ on that domain for which $ G_{Es}^{K^\ast}<0$.
(N.B.\ Here, analysed in terms of a common definition of $x=Q^2/m_{K^\ast}^2$, the light-quark zero is closer to $x=0$ than that for the heavier-quark.  This seeming conflict with the preceding discussion is explained below.)

The radii of the flavour-separated $K^\ast$-meson electric form factors can readily be computed from the results in Table~\ref{static}:
\begin{subequations}
\begin{align}
(r_{Eu}^{K^\ast})^2 & = r_{E}^{K^{\ast +}} - r_{E}^{K^{\ast 0}}  \phantom{2} = (0.70\,{\rm fm})^2 ,
\\
(r_{Es}^{K^\ast})^2 & = r_{E}^{K^{\ast +}} + 2 r_{E}^{K^{\ast 0}} =  (0.52\,{\rm fm})^2.
\end{align}
\end{subequations}
(Analogous formulae are valid for all form factors.)  Unsurprisingly, that associated with the heavier $s$-quark is smallest.
Analogous results for the magnetic moments and radii, and quadrupole moments are:
\begin{equation}
\label{flavoursymmetryV}
\begin{array}{l|ccc}
   & \mu_{K^\ast} & r_{K^\ast}^\mu/{\rm fm} & Q_{K^\ast}\\ \hline
u & 2.48 & 0.66 & -0.29\\
s & 1.70 & 0.51 & -0.35
\end{array} .
\end{equation}
Plainly, there are differences between the charge and magnetisation distributions of $u$- and $s$-quarks within the $K^\ast$; and also the associated quadrupole deformations.
($|Q_{K^\ast}^s|>|Q_{K^\ast}^u|$ because $|F_{2u}^{K^\ast} - F_{2s}^{K^\ast}|>|F_{3u}^{K^\ast} - F_{3s}^{K^\ast}|$, \emph{i.e}.\ the difference between $u$- and $s$-quark magnetic moments is large.  This is also found elsewhere \cite{Bhagwat:2006pu}.)
As with kindred features in the pseudoscalar meson sector \cite{Shi:2015esa, Chen:2016sno, Gao:2017mmp}, the size of such SU$(3)$ flavour-symmetry breaking effects in vector mesons, $28(9)$\%, is  determined by nonperturbative dynamics; namely, the current-quark-mass dependence of DCSB.

\begin{figure}[t]
\includegraphics[width=0.4\textwidth]{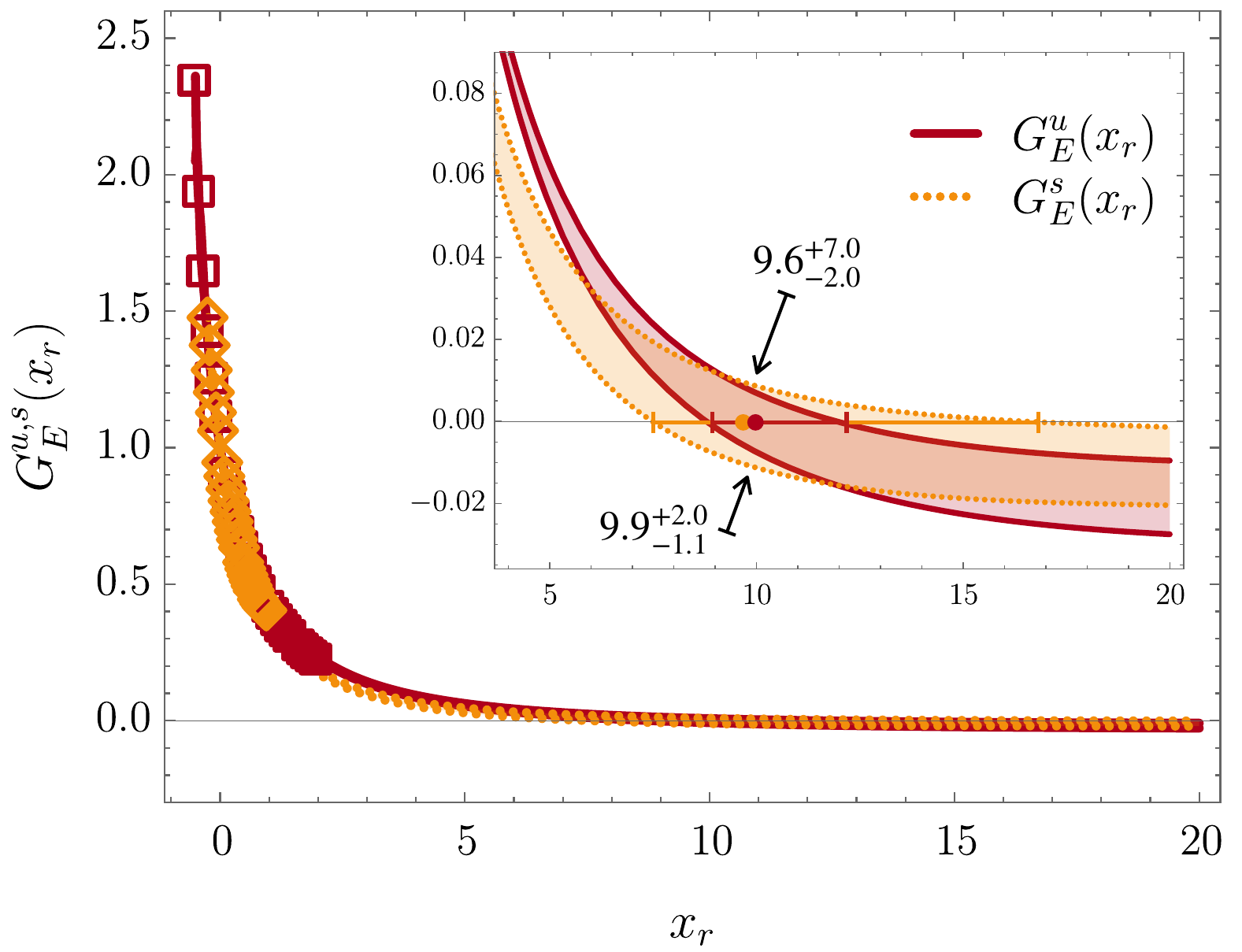}
\caption{\label{rescale}
Flavour-separated $K^\ast$ electric form factors.
$G_{Eu}^{K^\ast}$ is plotted versus $x_r=^\prime = Q^2/m_\rho^2$ -- solid (red) curve marked by squares; and
$G_{Es}^{K^\ast}$ versus $x_r=x^{\prime\prime} = Q^2/m_\phi^2$ -- dotted (orange) curve, diamonds.
(\emph{NB}.\ $m_\phi = m_{\rho_s}$.)
}
\end{figure}

Whilst not immediately apparent, the location of the zeros in $G_{Eu}^{K^\ast}$ and $G_{Es}^{K^\ast}$ can also be understood using the arguments developed after Eq.\,\eqref{GEzero}.
To elucidate, we note that $G_E^{K^\ast}(x_{K^\ast}=Q^2/m_{K^\ast}^2)$ does \emph{not} exhibit a pole at $x=-1$ because a virtual photon cannot transition to a $K^\ast$-meson.  Instead, it has two poles, \emph{viz}.\ one at $x_u^{K^\ast} = Q^2/m_\rho^2 = -1$, generated in the dressed--photon--$u$-quark vertex, and the second at $x_s^{K^\ast} = Q^2/m_\phi^2 = -1$, arising from the photon--$s$-quark vertex.  We make this explicit via Fig.\,\ref{rescale}, which depicts $G_{Eu}^{K^\ast}(x^\prime = x_{K^\ast}\, m_{K^\ast}^2/m_\rho^2)$ and $G_{Es}^{K^\ast}(x^{\prime\prime} = x_{K^\ast} m_{K^\ast}^2/m_{\phi^2})$.  Analysed this way, the curves indicate that the electric form factor of a heavier quark possesses a zero closer to the appropriate $x$-axis origin than does that of a lighter quark within the same bound state.  Owing to the breadth of the error band surrounding the $\bar s$-quark curve, one cannot be more categorial than this; but it is plain upon comparison with Fig.\,\ref{FigGEKstFlavour} that both zeros have moved a long way toward reordering.

\begin{figure*}[t]
\includegraphics[width=1.0\textwidth]{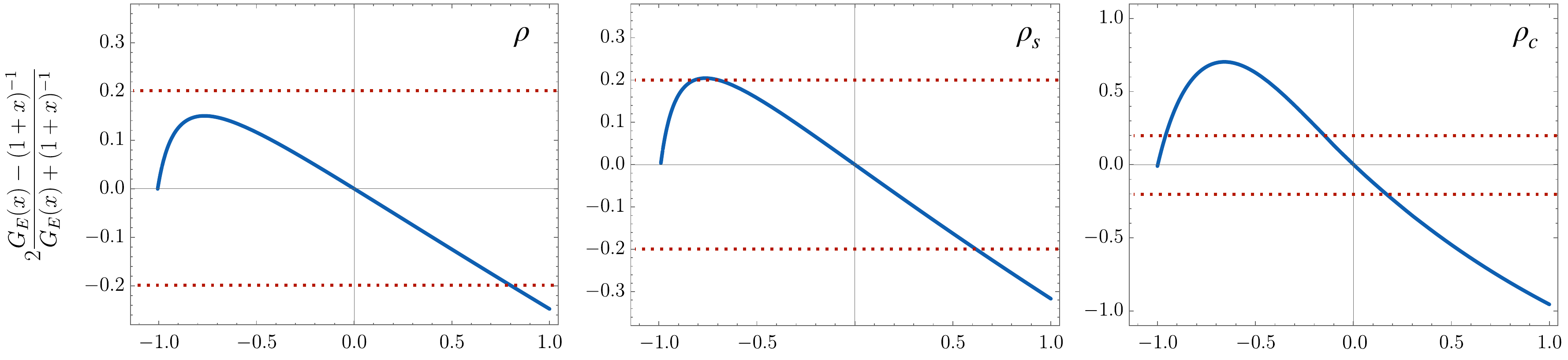}
\caption{\label{vmdfail}
Relative-discrepancy ratio in Eq.\,\eqref{VMDoff}, which measures the accuracy of a single-pole vector meson dominance approximation for $G_E^{\mathpzc V}(x)$, ${\mathpzc V}= \rho$, $\rho_s$, $\rho_c$.  The horizontal dotted lines bound the $\pm 20$\% range.
}
\end{figure*}

\subsection{Vector Meson Dominance}
The existence of a zero in vector meson form factors has another important corollary; namely, single-pole vector-meson-dominance (VMD), viz.\ $G_E^{\mathpzc V}(x) \approx 1/(1-x)$, can only be a useful tool for approximating (off-shell) vector meson properties within a limited $x$-domain.  We have analysed this; and in Fig.\,\ref{vmdfail} depict a discrepancy ratio:
\begin{equation}
\label{VMDoff}
\delta^{\mathpzc V}(x) := 2 \frac{G_E^{\mathpzc V}(x) - 1/(1-x)}{G_E^{\mathpzc V}(x) + 1/(1-x)},
\end{equation}
with $x=Q^2/m_{\mathpzc V}^2$ for ${\mathpzc V}= \rho$, $\rho_s$, $\rho_c$.

The vector-meson electric form factor presents the best case for a VMD model because it necessarily agrees with the computed result in some neighbourhood of $x=-1$ and, by charge conservation, also in the vicinity of $x=0$.  Our analysis reveals that the discrepancy is less-than 20\% within the following regions:
\begin{equation}
\label{VMDregions}
\begin{array}{lccc}
\rho_{\phantom{s}}\,: &-1 < x < 0.81\,, & & \\
\rho_s\,: &-1 < x < 0.60\,, & & \\
\rho_c\,: &\; -1 < x < -0.96 & \mbox{\&} & -0.15 < x < 0.24\,.
\end{array}
\end{equation}
(We do not consider $x<-1$.)
%
%
It is clear from Eq.\,\eqref{VMDregions} and Fig.\,\eqref{vmdfail} that a single-pole VMD approximation is a fair assumption on a reasonable domain for light-quark systems.  However, it is poor for states in which the Higgs-mechanism of mass generation is dominant, \emph{i.e}.\ $c \bar c $ and more massive systems.  In fact, without the $x=0$ constraint imposed by current conservation, a VMD approximation for the $c \bar c $ system becomes quantitatively unreliable once bound-state virtuality exceeds 4\%.

\section{Summary and Perspective}
\label{epilogue}
The symmetry-preserving rainbow-ladder truncation of QCD's continuum bound-state equations was used to calculate the elastic electromagnetic form factors of the $\rho$-meson and subsequently study the evolution of such vector-meson form factors with current-quark mass.  In addition, to enable a full comparison with kindred treatments of pseudoscalar meson form factors and explore the environmental sensitivity of quark contributions, $K^\ast$-meson elastic form factors were also computed.

Predictions for the static properties of these systems are listed in Table~\,\ref{static}, which reveals that electric charge radii of vector mesons are typically $\sim 10$\% larger than those of the related pseudoscalar mesons.  Importantly, the product of vector-meson charge-radius and leptonic decay constant is practically constant on the domain of meson masses within which emergent mass is dominant [Eq.\,\eqref{ftimesr}].  This matches the behaviour in the pseudoscalar sector, albeit therein the analogous product is $\approx 40$\% smaller.  Evidently, emergent mass also plays a prominent role in fixing vector meson properties.
Furthermore, a simultaneous analysis of the magnetic and quadrupole form factors on $Q^2\simeq 0$ shows significant deformation of each vector meson: relative to point-particle values, the magnetic and quadrupole moments deviate by $33(7)$\% [Eq.\,\eqref{deformation}].  Notably, over a $250$-fold increase in current-quark mass, from $\hat m_u \to \hat m_c$, these quantities are practically unchanged.

The comparison between $\rho$- and $K^\ast$-meson elastic form factors exposed additional similarities with the pseudoscalar meson sector, \emph{e.g}.\ as with an array of $K$-meson properties, the magnitude of $SU(3)$-flavour-symmetry violation in $K^\ast$ mesons is commensurate with the value of $f_{K^\ast}/f_\rho$ [Eq.\,\eqref{flavoursymmetryV}]; namely, it is set by the flavour-dependence of dynamical chiral symmetry breaking (DCSB).

Experimental data from JLab, which suggest that the proton's electric form factor, $G_E^p$, might pass through zero at $Q^2/m_p^2 \approx 10$, focus attention on vector meson electric form factors because, like $G_E^p$, the vector-meson electric form factor, $G_E^{\mathpzc V}$, is a sum of terms, one of which is negative-definite.  Hence, studies of $G_E^{\mathpzc V}$ may provide qualitatively sound guidance on the possible appearance and location of a zero in $G_E^p$.  This capacity is especially useful because the meson bound-state problem is more easily solved than that for the baryon.  It was found herein that $G_E^{\mathpzc V}$ always exhibits a zero; and analysed as a function of $x=Q^2/m_{\mathpzc V}^2$, that zero moves toward $x=0$ with increasing current-quark mass [Eq.\,\eqref{GEzero}].  These features can also be understood as consequences of DCSB; and they support a qualitative argument that the character of emergent mass in the Standard Model may ensure a zero in $G_E^p$.

The existence of a zero in $G_E^{\mathpzc V}(x)$ entails that the domain within which a single-pole vector meson dominance model can serve as a useful approximation to vector meson properties is circumscribed.  Notwithstanding this, $G_E^{\mathpzc V}(x)$ is the best case for a VMD model because it must agree with the computed result in some neighbourhood of $x=-1$ and, by charge conservation, also in the vicinity of $x=0$.  It was found herein that a single-pole VMD approximation is a fair assumption on a reasonable domain for light-quark systems [Eq.\,\eqref{VMDregions}].  However, it is poor for states in which the Higgs-mechanism of mass generation is dominant; hence, it is likely to yield erroneous estimates for the off-shell properties of $c \bar c $ and more massive systems.

Focusing on the $Q^2$-dependence of form factors, a particular feature of the analysis herein was use of the Schlessinger point method (SPM) for the continuation of results into a $Q^2$-domain that was not accessible using brute-force numerical techniques.  As a growing number of applications have shown, the SPM is remarkably reliable.  Its power opens the way for numerous extensions of this study; \emph{e.g}.\ analyses of the semileptonic decays of $D$, $D_s$ mesons are underway and, naturally, of the proton and neutron elastic electromagnetic form factors.

\acknowledgments
We are grateful for constructive comments and encouragement from Y.~Lu, S.-X.~Qin and J.~Segovia.
Work supported by:
Innovation Program of Jiangsu Province;
Jiangsu Province \emph{Hundred Talents Plan for Professionals};
National Natural Science Foundation of China, under grant nos.~11805097, 11847024, 11905107, and 11535005;
Jiangsu Province Natural Science Foundation, under grant nos.~BK20180323 and BK20190721;
Nanjing University of Posts and Telecommunications Science Foundation, under grant no.~NY129032;
and
Natural Science Foundation of the Jiangsu Higher Education Institutions of China 19KJB140016.



\end{document}